# Lightning as a natural accelerator of relativistic particles and a source of synchrotron radiation

N.I. Petrov

Scientific and Technological Centre of Unique Instrumentation of the Russian Academy of Sciences, 15 Butlerova str., Moscow, 117342 Russia

E-mail: petrovni@mail.ru

***Abstract***

A physical model of the formation of a vortex current in a plasma channel of lightning discharges due to the anomalous Hall effect caused by relativistic spin-orbit interaction phenomenon is proposed. The existence of fast electromagnetic surface plasmon waves propagating along the lightning discharge channel at a speed close to the speed of light in vacuum is shown. The spin properties of electrons become important at relativistic velocities, so the spin-orbit coupling arises as an effective magnetic force acting on the spin of an electron, leading to the formation of a spin-induced vortex current in the absence of an external magnetic field. This makes it possible to consider the lightning discharge channel as a charged particle accelerator that generates microwave, X-ray, gamma-ray and RF emissions by a current pulse moving along a helical trajectory via synchrotron and cyclotron radiation mechanisms. The proposed theoretical model of the formation of a spin-induced vortex current in the lightning plasma channel is consistent with observations of lightning and spark discharges.

## Introduction

The current flow in classical plasmas is described in terms of the number of charge carriers-electrons and their speed. However, electrons are the particles that have a magnetic moment (spin) along with a charge. Recently, the integration of quantum effects into the dynamics of classical plasma has been of great interest [1-3]. It was shown in [2] that the spin properties of electrons can be important even in a high-temperature plasma with a modest density and magnetic field strength. In classical plasma, spin effects can be neglected due to the random thermal orientation of the spin vector. However, under specific conditions, they cannot be neglected, and they can be noticeable. We emphasize here, that the spin-induced forces acting on electrons in a plasma moving with relativistic velocities become significant. In particular, the spin properties of electrons become important at the charge neutralization stages of a lightning when the relativistic effects are amplified due to the achievement of almost light velocities. It was established that the front of the charge neutralization moves along the lightning channel at a speed of the order of the speed of light [4]. Note that the propagation of the fast ionization waves or ionizing waves of potential gradient under electrical breakdown conditions were also experimentally demonstrated in [5, 6]. It was shown that the shorter the pulse front of the applied voltage and the stronger the pre-ionization of the discharge gap, the greater the speed of the ionization wave starting from the high-voltage electrode. The measured speeds of the return stroke and illumination pulses are less than the speed of light [7-9]. Optical measurements of return strokes speed are not available during the initial stages of natural lightning return strokes, but it can be evaluated from the measured electric fields and electric field derivatives. It was shown in [10] that the initial return stroke speed in the triggered lightning channel is near the speed of light.

In recent decades, a phenomenon of terrestrial gamma-ray flashes (TGF) generated during thunderstorms was discovered [11]. X-ray and gamma-ray flashes from natural [12, 13] and rocket-triggered [14] lightning, as well as from laboratory spark discharge were observed [15]. TGF photons are usually assumed to be produced via the bremsstrahlung of runaway electrons accelerated by strong electric fields in the atmosphere [16]. Currently, two models of electron acceleration and multiplication based on the relativistic runaway electron avalanche (RREA) mechanism [4, 17] and the lightning leader model [18, 19] have been proposed. These models explain the multiplication of energetic electrons and the subsequent production of bremsstrahlung photons. However, some basic properties, such as the radiation spectrum (in the region above 10 MeV), the cut-off energy, and the polarization and beaming characteristics of the radiation still need to be explained. Clarifying the relationship between the parameters of TGF and the characteristics of lightning discharge channel is also important. From the ground-based observations, it follows that TGFs are associated with cloud-to-ground discharges of negative polarity. The measurements show that X-rays from natural lightning and intense bursts of gamma-ray radiation with energies up to 10 MeV are correlated with negative leader stepping [4, 17]. The detection of TGF emission with photon energies in the 10-100 MeV range was reported in [20]. It was shown that the detected power-law radiation in the range from 10 to 100 MeV is difficult to be explained using RREA models [17-19]. The recent observations have shown that gamma radiation correlates with radio-frequency radiation and is generated at the last stage of lightning leader channel development prior to the lightning return stroke [21]. The bursts of 30-80 MHz radiation because of leader stepping were observed in [22]. It was shown in [23] that TGF is produced in the initial stage of a lightning flash just before the initiation of the current pulse. The observations of TGFs which occur at the onset of UV and optical emissions also point to the importance of lightning leaders [23-25].

In this paper, the propagation of the fast surface plasmon waves along a lightning discharge channel is investigated depending on the frequency and conductivity of the channel. The physical mechanism of the formation of the vortex current flow along a spiral plasma channel in the absence of an external magnetic field is proposed. It is emphasized that the vortex current is a consequence of the influence of an effective magnetic field arising due to the relativistic spin-orbit interaction. It is shown that the current pulse moving along the spiral trajectory with relativistic speed can produce microwave, THz, X-ray, and gamma-ray emissions via cyclotron and synchrotron radiation mechanisms.

## 1. Fast surface plasmon waves propagating along the lightning channel

The streamer-leader process underlies the development of lightning and spark discharges in the atmosphere [4]. The embedded charges distributed within the leader channel are neutralized during the leader step processes or during the lightning return stroke. It is established that the front of the neutralization process moves along the channel at a speed of the order of the speed of light. High frequency electromagnetic wave modes are excited in the lightning discharge channels during the leader step processes or during grounding. It is well known that the surface electromagnetic waves can propagate along the conducting wire [26].

### 1.1. Velocity of surface waves

The electromagnetic field behaviour in a lightning channel is described by the dispersion equation which is followed from the Maxwell equations. For the cylindrical structure of plasma channel the guided modes can be determined from the Helmholtz equations for the longitudinal field component $E_z$:

$$\left[\nabla_\perp^2 + \left(k_0^2\varepsilon_p - \beta^2\right)\right]E_z = 0\,, \quad 0 < r < r_0$$
$$[\nabla_\perp^2 + (k_0^2 - \beta^2)]E_z = 0, \quad r > r_0, \tag{1}$$

where $\nabla_\perp^2 = \frac{1}{r}\frac{\partial}{\partial r}\left(r\frac{\partial}{\partial r}\right) + \frac{1}{r^2}\frac{\partial^2}{\partial\varphi^2}$, $k_0 = \frac{\omega}{c}$ is the wavenumber in free space, $\beta$ is the longitudinal component of the wavenumber, $r_0$ is the channel radius, $\varepsilon_p = \varepsilon' + i\frac{\sigma}{\omega\varepsilon_0}$ is the complex dielectric constant, where $\sigma = \frac{1}{R_l \pi r_0^2}$ is the electric conductivity, $R_l$ is the resistance per unit length, and $\varepsilon_0$ is the dielectric constant of free space.

Solutions of the Eqs. (1) are the Bessel functions:

$$E_z = \begin{cases} A_1 I_0(\eta r), r \leq r_0 \\ A_2 K_0(\eta_0 r), r \geq r_0 \end{cases}, \tag{2}$$

where $A_1$ and $A_2$ are the amplitude coefficients, $I_0$ and $K_0$ are the modified Bessel functions of the first and second kind, $\eta^2 = \left(\frac{\omega^2}{c^2}\right) \varepsilon_p - \beta^2$, and $\eta_0^2 = \left(\frac{\omega^2}{c^2}\right) - \beta^2$.

The wave fields (2) are localized near the plasma-air boundary, so the wave is a surface wave propagating along the channel boundary.

The dispersion equation for the surface electromagnetic waves is followed from the boundary condition of continuity of the tangential components of the field at $r = r_0$:

$$\frac{\varepsilon_p}{\eta a} \frac{I_0'(\eta a)}{I_0(\eta a)} = \frac{1}{\eta_0 a} \frac{K_0'(\eta_0 a)}{K_0(\eta_0 a)}, \tag{3}$$

where $I_0$ and $K_0$ are the modified Bessel functions of the first and second kind, $I_0^{'}$ and $K_0'$ are the derivatives of the Bessel functions, $\eta^2 = k_0^2 \varepsilon_p - \beta^2$, $\eta_0^2 = k_0^2 - \beta^2$, $k_0 = \frac{\omega}{c}$ is the wavenumber in free space, $\beta$ is the longitudinal component of the wavenumber, $r_0$ is the channel radius, $\varepsilon_p = \varepsilon^{'} + i\frac{\sigma}{\omega \varepsilon_0}$ is the complex dielectric constant, where $\sigma = \frac{1}{R_l \pi r_0^2}$ is the electric conductivity, $R_l$ is the resistance per unit length, and $\varepsilon_0$ is the dielectric constant of free space.

The phase and group velocities of the surface wave can be determined from the dispersion Eq. (3). The wavevector $\beta = \beta^{'} + i\beta^{''}$ is a complex value. The real part $\beta'$ defines the phase velocity $V_{p□} = \frac{\omega}{\beta^{'}}$ of the wave, and the group velocity is determined by $V_g = \frac{d\omega}{d\beta'}$. The imaginary part $\beta^{''}$ defines the attenuation length $z_0 = \frac{1}{\beta^{''}}$ of the surface wave propagating along the discharge channel.

In Fig.1, the velocity and the attenuation length of the surface wave as a function of the frequency, conductivity and radius of the discharge channel are presented. It follows from the simulation that the velocity increases with the frequency. The propagation distance of the surface wave decreases when the frequency increases (Figs. 1a and 1b). Only the components of the current pulse at the frequencies of the order of MHz and lower reach high altitudes of the order of a kilometer or more. At a given frequency, the velocity is higher at the higher electric conductivity of a channel and the propagation distance increases with the conductivity (Figs. 1c and 1d). The velocity and propagation distance of the surface waves increase with the discharge channel radius (Figs. 1e and 1f).

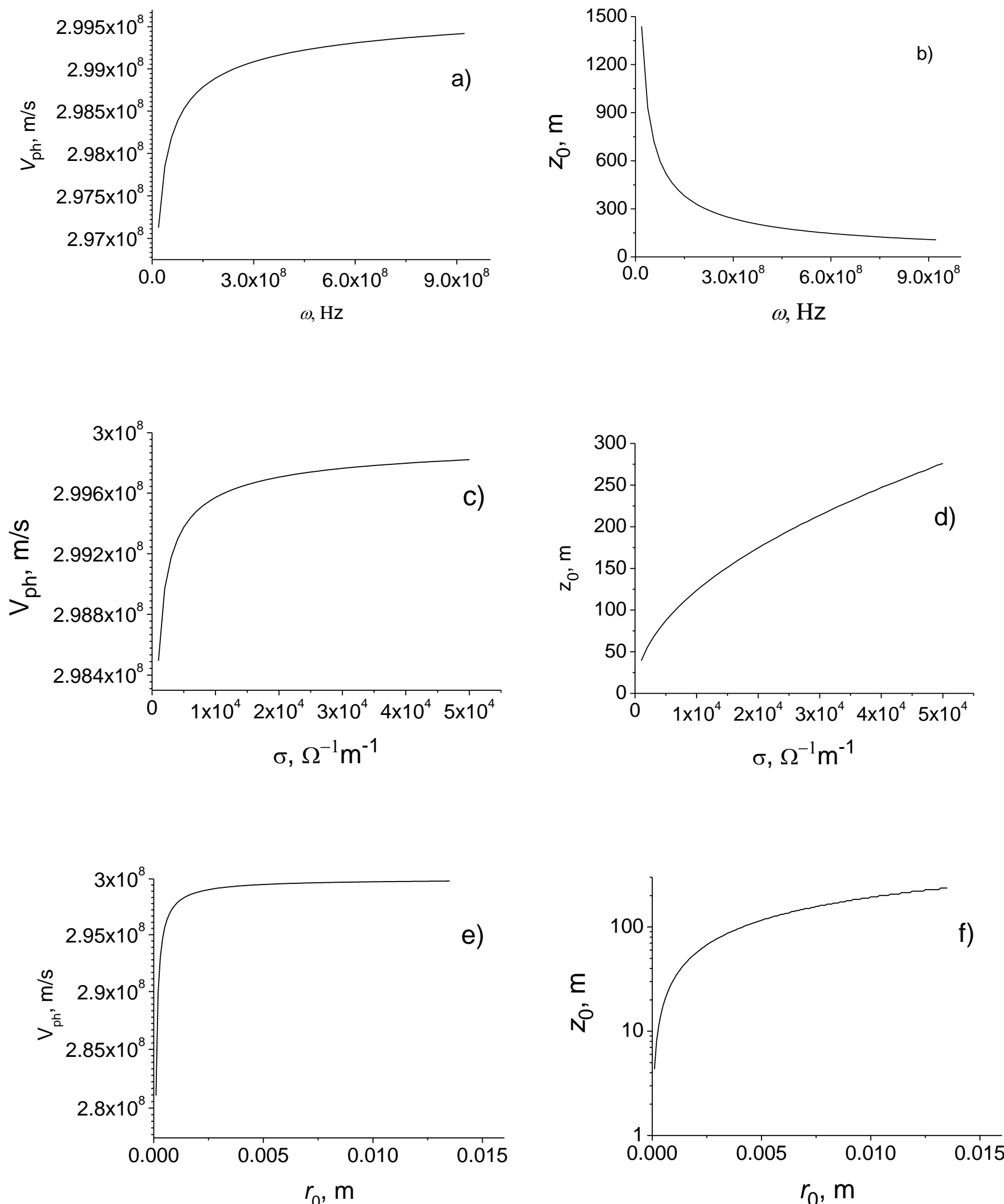

**Fig. 1.** Velocity (panels a, c, and e) and attenuation length (panels b, d and f) as function of frequency (a, b), conductivity (c, d) and channel radius (e, f). (a, b) - $\sigma = 10^4\,\Omega^{-1}m^{-1}$, $r_0 = 0.01$ m; (c, d) - $\omega = 2\pi f = 10^9$ Hz, $r_0 = 0.01$ m; (e, f) - $\omega = 2\pi f = 3 \cdot 10^{10}$ Hz, $\sigma = 10^5\,\Omega^{-1}m^{-1}$.

Thus, the speed of surface waves increases with frequency, as well as with the conductivity and radius of a discharge channel (Fig. 1). These waves are attenuated during propagation along the channel because of the skin effect. The dissipation increases with frequency because of the skin effect. The propagation distance of surface waves decreases with frequency because of dissipation (Fig. 1b). However, this distance increases with the conductivity and radius of a discharge channel (Figs. 1d and 1e). It follows that the ultra-relativistic velocities of the order of the speed of light in vacuum are achieved for high-frequency waves at the conductivities and the radii of the discharge channel characteristic of lightning. The propagation distance of the surface wave decreases with the increasing frequency. As follows from the calculations, the group velocity is greater than the phase velocity due to abnormal dispersion in the plasma channel. In a medium without dispersion, the group velocity of propagating modes is less than the phase velocity [27].

Note that the surface plasmon wave causes an additional ionization resulting in the generation of the higher plasma density. It was shown that the electron density of the lightning stepped leader is of the order of $10^{24}$ m$^{-3}$ [28]. The measurements show that the resistances per unit length of the lightning channel are of the order of $10^{-2}$–$10^{-1}$ Ω/m and the internal electric field strengths are of the order of $10^{3}$ V/m [29]. For the resistance per unit length $R_l = 10^{-2}$ Ω/m and the channel radius $r_0$ = 6 mm we have the conductivity $\sigma = \frac{1}{R_l \pi r_0^2} \approx \frac{1}{10^{-2} \cdot 10^{-4}} \sim 10^6$ $\Omega^{-1}$m$^{-1}$. The same order of conductivity follows from the Drude model for the electron density of the lightning leader $n = 10^{24}$ m$^{-3}$. In the laboratory spark discharges, the conductivity has lower values: $R_l \sim 10$ Ω/m [30, 31].

Measurements show that the optical return stroke speed (the optical radiation wave propagation speed $v_{opt}$) in a negative lightning is between $0.6 \cdot 10^8$ m/s and $2 \cdot 10^8$ m/s [7-9]. There is a time delay between the return stroke current and the luminosity, i.e., the current wave velocity $v_{cur} > v_{opt}$. For positive polarity lightning the return stroke speeds of the order of $10^8$ m/s are recorded. It was shown in [10] that the initial return stroke speed is near the speed of light $c$ for the bottom 30 m of the triggered lightning channel. However, these speeds do not take into account the tortuosity of the channel. When considering the small-scale tortuosity of the current trajectory, these velocities will be close to the speed of light in a vacuum.

### 1.2. Radiation of e/m waves

Lightning radiation has a maximum intensity in the region of 5-20 kHz and its spectral density varies inversely with frequency [32]. The wavelength corresponding to the maximum radiation intensity has the order of the lightning channel length. The second characteristic size of the lightning channel is the characteristic tortuosity length of the channel (usually of the order of the streamer zone length of leaders), which corresponds to the frequency range of $f$ = 3 - 300 MHz (wavelength $\lambda$ = 1 - 100 m). There is another characteristic size — the transverse size of the lightning channel corresponding to the frequency range of $f$ = 300 - 3000 MHz ($\lambda$ = 0.1 - 1 m). If the frequency range of 5 - 100 kHz can be well described as the radiation of a linear dipole, then the observed high-frequency radiation cannot be explained in the dipole approximation. In recent decades, the phenomenon of terrestrial gamma-ray flashes generated during thunderstorms has been discovered. X-ray and gamma-ray flashes were observed from natural and rocket-launched lightning, as well as from laboratory spark discharge [17-25]. It is usually assumed that high-energy photons are formed due to the bremsstrahlung of runaway electrons accelerated by strong electric fields in the atmosphere [16]. Recent observations have shown that gamma radiation correlates with the radio frequency radiation [22]. In [33], we proposed a physical mechanism for the formation of ultrahigh-frequency (microwave) radiation in the spark discharge and lightning caused by a pulse of polarization current and the associated field of a surface plasmon wave.

An electromagnetic wave in a plasma channel results in a time-dependent electric dipole (polarization) that generates a polarization current pulse propagating at a relativistic speed. It follows from the laws of electrodynamics that charged particles or dipoles moving with acceleration (or oscillation) should emit electromagnetic radiation. A polarization current pulse moving along a curved (circular) path generates a radiation similar to the synchrotron radiation produced by electrons circulating in a magnetic field. Figure 2 shows a simplified model of induced polarization motion along the surface of a plasma channel boundary. Spatially decaying electric field causes the negative and positive charges to move in opposite directions, i.e., a polarization $P$ is induced. The motion of the polarized region at a speed $v \approx c$ generates the polarization current $j_{pol} = \frac{\partial P}{\partial t} = \frac{\partial P}{\partial z}\frac{\partial z}{\partial t} = \frac{1}{4\pi}(\varepsilon - 1)\frac{\partial E}{\partial z}v = (\varepsilon - 1)\rho v$, where $\rho = n_e - n_i$ is the charge density.

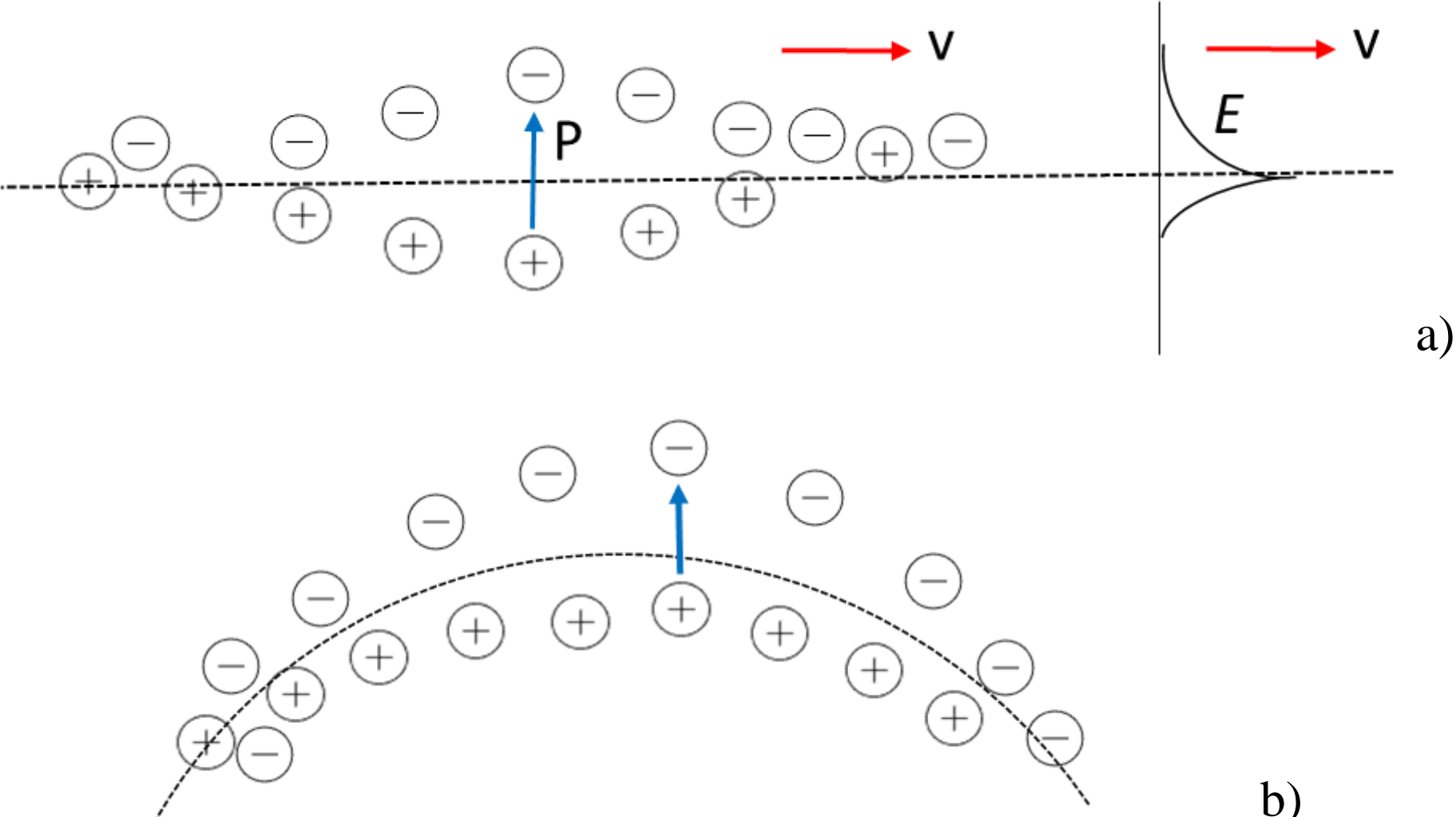

**Fig. 2.** Simplified representation of the motion of a polarized region. The velocity v coincides with the velocity of the surface electromagnetic field. In (a) on the right, the exponential dependence of the electromagnetic field intensity on the distance away from the interface is shown; (b) the channel curvature introduces centripetal acceleration of the moving polarized region and electrons.

The current density in the Maxwell equation for the electric field acquires an additional term:

$$\bar{J} = \sigma\bar{E} + \frac{\partial\bar{P}}{\partial t} = \sigma\bar{E} + \varepsilon_0(\varepsilon - 1)\frac{\partial\bar{E}}{\partial t}, \tag{4}$$

where $\bar{J}_{cond} = \sigma\bar{E}$ is the conduction current and $\bar{J}_{pol} = \frac{\partial\bar{P}}{\partial t}$ is the polarization current.

The induced polarization defines the electric dipoles density, so the properties of the radiation are determined by the emission resulting from the motion of a dipole or from a charge $e$ that moves with the relativistic speed along a circular trajectory. The contribution of ions to the plasma polarization is only a small correction of the order of the ratio of the masses of electrons and ions $m_e/M_i \sim 10^{-3}$, so it can be neglected. Note that the velocity of the electrons involved in the conduction current is much less than the velocity of the polarization waves. However, the polarization induced by the surface electromagnetic field propagates along the lightning leader channel with an ultra-relativistic velocity at relatively high conductivities of the lightning channel.

## 2. Spin-induced vortex current formation

As was shown above, the return stroke current pulses in lightning propagate with the relativistic velocities close to the speed of the light in vacuum. Therefore, it is advisable to use the Dirac equations to describe the behavior of the particles in a lightning discharge channel. Scattering of the unpolarized electrons by an unpolarized target results in spatial separation of the electrons with different spins due to spin-orbit interaction. In the non-relativistic approximation, the electrical interaction of the charged particles does not depend on their spins because the electrical interaction exceeds the spin-orbit interaction. However, when relativistic effects are considered, this interaction becomes spin dependent. In this case, the Hamiltonian describing an electron in an external field has the form [34]:

$$\hat{H} = \frac{\hat{p}^2}{2m} + e\Phi - \frac{\hat{p}^4}{8m^2c^2} - \frac{eh}{2m^2c^2}\hat{s}\left(\vec{E} \times \vec{p}\right) - \frac{e^2h^2}{8m^2c^2}div\bar{E}, \tag{5}$$

where $\vec{E} = -\nabla\Phi$ is the electric field, $e$ is the electron charge, $m$ is the electron mass, ħ is the Planck constant, $c$ is the speed of light, $\hat{s}$ is the operator of the electron spin, and $\hat{p} = -i\hbar\nabla$ is the momentum operator.

The last three terms in Eq. (5) describe relativistic corrections of the order of $\left(\frac{v}{c}\right)^2$ to $\frac{\hat{p}^2}{2m} + e\Phi$. The first of them is a consequence of the relativistic dependence of the kinetic energy on momentum. The second one is the interaction energy of the moving magnetic moment with the electric field, i.e., the energy of the spin-orbit interaction. The last term is different from zero only at those points where there are charges that create an external field.

Energy of spin-orbit interaction is

$$\mathcal{E}_m = -\frac{e\hbar}{2m^2c^2}\bar{s}(\bar{E}\times\bar{p}) = -\bar{\mu}\bar{B}_{eff} = -\frac{1}{2c^2}\bar{\mu}(\bar{E}\times\bar{v}) ,$$

where $\vec{\mu} = \frac{e\hbar}{m}\vec{s}$ is the magnetic moment of the electron, and $\bar{B}_{eff} = \frac{1}{2mc^2}(\bar{E}\times\bar{p})$ is the effective magnetic field acting on the electron.

The spin-orbit interaction term in Eq. (5) appears as an effective magnetic force acting on the electron spin. The transverse spin-orbital force is given by

$$F_{s0} = \nabla\mathcal{E}_m = \frac{e\hbar\nabla}{4m^2c^2}[s(\nabla\Phi\times p)] \sim eE\frac{p^2}{m^2c^2}$$

Thus, the spin-orbital force has the order of the electric force for relativistic electrons, i.e. at $v \to c$. An effective magnetic force acts on the electron spin leading to the formation of a spin-induced vortex current flow in the absence of an external magnetic field. If the electric field is centrally symmetric, then $\boldsymbol{E} = -\frac{\boldsymbol{r}}{r}\frac{d\Phi}{dr}$, where for the radial electric field created by the cylindrical lightning channel with the radius $r_0$ and the charge density per unit length $q$ we have

$$E_r = \frac{q}{2\pi\varepsilon_0 r_0} .$$

It follows from the measurements that for the conditions at sea level, the electric field in the streamer zone of the positive leader is about 5 kV/cm, and the electric field in the streamer zone of the negative leader is about 10 kV/cm [35]. Applying the Gauss theorem to estimate the charge density per unit length, for a radial electric field we obtain

$$E = \frac{q}{2\pi\varepsilon_0 r_0} \approx 10^7 \text{ V/m}. \tag{6}$$

Hence, for the effective magnetic field we have

$$B_z^{eff} = \frac{1}{2c^2}E_r v_\varphi \approx \frac{10^7\cdot 2\cdot 10^8}{2\cdot 9\cdot 10^{16}} \sim 10^{-2}T . \tag{7}$$

In a homogeneous plasma, the generalized Ohm's law with the Hall parameter $\Omega \cong \omega_c\tau_e$ is fulfilled for the current density:

$$\boldsymbol{J} = D\nabla n + \sigma\vec{\boldsymbol{E}} + \sigma\left(\bar{\boldsymbol{v}}\times\vec{\boldsymbol{B}}_{eff}\right) = D\nabla n + \vec{\boldsymbol{j}} + \vec{\boldsymbol{j}}\times\boldsymbol{\Omega}, \tag{8}$$

where $\boldsymbol{j} = -en\boldsymbol{v}_e$, $\boldsymbol{\omega}_c = e\boldsymbol{B}/m_e$, $\sigma = e_0^2 n\tau_e/m_e$, $\tau_e$ is the electron attachment time, $\boldsymbol{v}$ is the charge velocity associated with the surface plasmon-polariton wave, *D* is the diffusion coefficient, ***B*** is the magnetic field, and ***E*** is the electric field.

The cyclotron frequency is equal to $\omega_c = \frac{eB_{eff}}{m} \approx \frac{1.6 \cdot 10^{-19} \cdot 10^{-2}}{10^{-30}} \approx 10^9\, s^{-1}$, hence for $\nu_e \approx 10^7\, s^{-1}$, the Hall parameter $\Omega = \frac{\omega_c}{\nu_e} = \omega_c \tau_e = \frac{eB_{eff}}{m\nu_e} \gg 1$.

The first two terms in Eq. (8) describe dissipative effects, and the last term describes non-dissipative effects. This indicates that the Hall current does not affect the Joule losses.

In a plane that is perpendicular to a magnetic field a charge particle experiences the centripetal acceleration. This means that the charge undergoes cyclotron motion. The radius of the cyclotron motion can be estimated from the Newton's second law $F = ma_r = mv^2/R$. Equating this force to the Lorentz force $F_L = evB_{eff}$ we have

$$R = \frac{mv}{eB_{eff}} \approx \frac{10^{-30} \cdot 2 \cdot 10^{8}}{1.6 \cdot 10^{-19} \cdot 10^{-2}} \sim 10^{-1}\ \text{m}.$$

Thus, the radius of the cyclotron motion is of the order of the channel radius: $R \sim r_0$.

The magnitude of the vortex current is determined by the radius of cyclotron motion *R* and the value of the magnetic field:

$$i_v = \frac{2BR}{\mu_0} \approx \frac{2 \cdot 10^{-2} \cdot 10^{-1}}{1.3 \cdot 10^{-6}} \sim 1.5 \cdot 10^{3}\ \text{A},$$

where $\mu_0$ is the permeability of free space.

It follows that the vortex current is one or two orders of magnitude less than the peak values of the return stroke current [4, 30, 31].

## 3. Synchrotron radiation from lightning

In a magnetic field, an electron moves along a helical trajectory with an axis on a magnetic field line. At the same time, it experiences acceleration, which leads to radiation. The mechanism of the radiation of a charged particle in a magnetic field is generally called the magnetic bremsstrahlung. The radiation of a nonrelativistic particle ($E \ll mc^2$), usually called cyclotron, for the relativistic particles is a synchrotron.

When magnetic fields are present, the charges can interact with them and radiate or absorb the radiation. For the slowly moving particles, this happens at a single frequency - the cyclotron frequency. For particles moving at relativistic speeds, radiation or absorption occurs over a wide frequency range and in this case is called synchrotron radiation.

There are numerous curvatures and irregularities on the tortuous lightning channel boundary (Fig. 3a). The curvature of the trajectory introduces centripetal acceleration in the moving polarized region, thereby leading to electromagnetic radiation. Acceleration and radiation also occur when electrons move along a spiral trajectory (Figs. 3b and 3c).

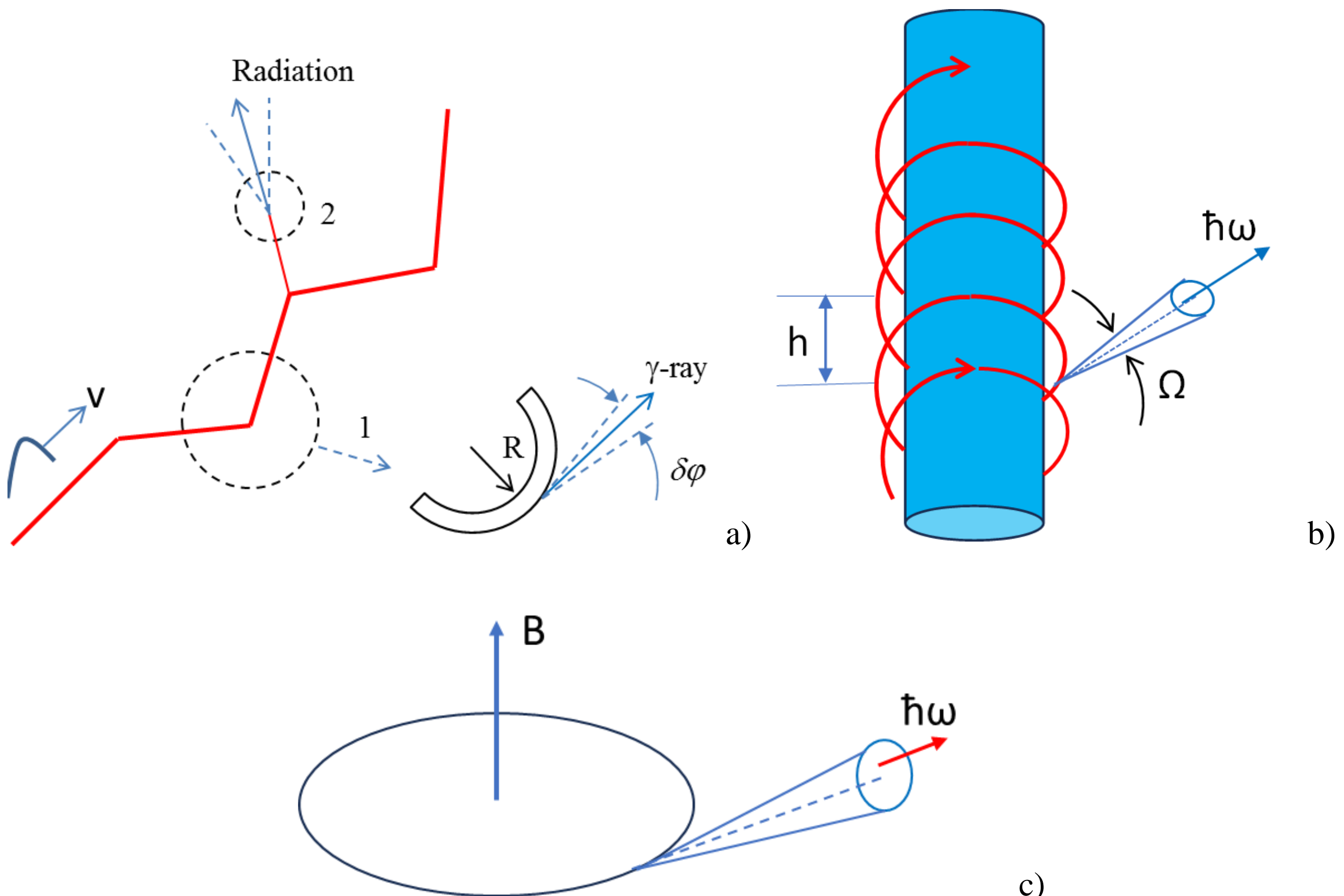

**Fig. 3.** Schematic view of a lightning channel tortuosity. (a) Sources (places) of radiation: 1 – synchrotron radiation, 2 – transition radiation; (b) lightning channel (blue color) and vortex current (red color), (c) B – magnetic field.

Electromagnetic radiation generated by charges moving along a curved trajectory (synchrotron radiation) has been known for a long time [36].

The power emitted into the $m$-th harmonic is given by [37]:

$$W(m) = \frac{e^2 cm\alpha}{4\pi\varepsilon_0 R^2}\left[2\alpha^2 J_{2m}^{'}(2m\alpha) + (1-\alpha^2)\int_0^{2m\alpha} J_{2m}(x)dx\right], \tag{9}$$

where $\alpha = \frac{v}{c}$, $m$ is the harmonic order, $R$ is the curvature radius of the trajectory bend.

In the non-relativistic case, the main contribution to the total power $W = \sum_{m=1}^{\infty} W(m)$ gives the radiation of the first harmonic ($m$ = 1, dipole radiation). The total radiation power of a non-relativistic electron is [37]:

$$W(1) = \frac{2}{3}\frac{e^2 c\alpha^4}{4\pi\varepsilon_0 R^2}. \tag{10}$$

In the ultra-relativistic case, the total radiation power is given by [37]:

$$W = \frac{2}{3}\frac{e^2 c}{4\pi\varepsilon_0}\frac{\alpha^4\gamma^4}{R^2}, \tag{11}$$

where $\gamma = \left(1 - \frac{v^2}{c^2}\right)^{-\frac{1}{2}}$ is the relativistic Lorentz factor, $\alpha = \frac{v}{c}$, $e$ is the electron charge, v is the velocity of the surface plasmon wave (polarization density), $c$ is the speed of light in vacuum, and $R$ is the curvature radius of the trajectory bend.

In Fig. 4, the photon frequency $\omega_{ph} = E_{ph}/\hbar$ and photon energy $E_{ph} = W/\omega_0$ as a function of the conductivity are shown. The power $W$ is determined from Eq. (11). The values of γ and α in Eq. (11) for a given conductivity and the radius of the lightning channel are determined from the solution of Eq. (3).

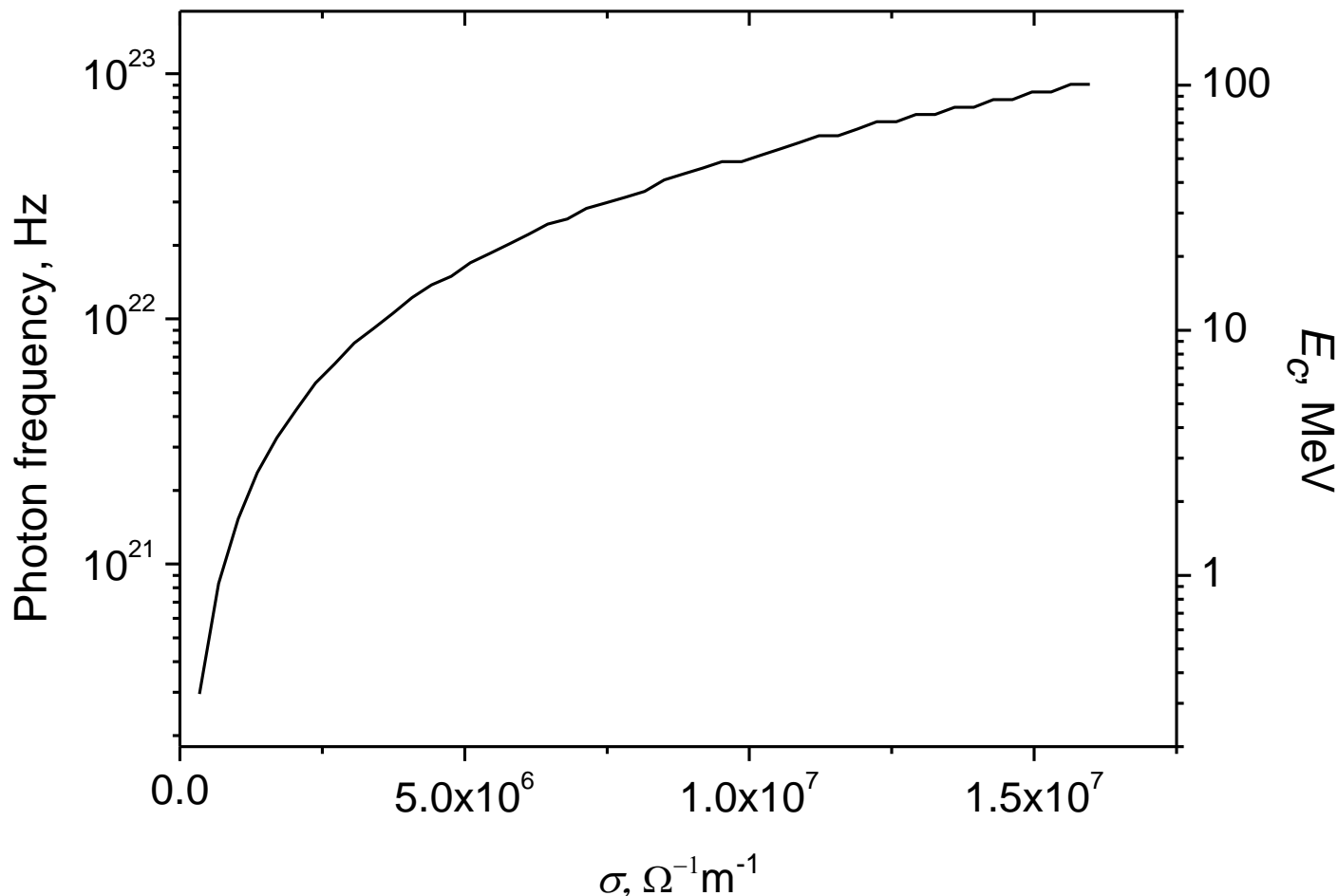

**Fig. 4.** The photon frequency and energy as function of conductivity. $\omega = 2\pi f = 3.1 \cdot 10^{10}$ Hz, $r_0$ = 0.01 m.

### 3.1. Radiation power spectrum

Power spectrum is defined by [37]

$$\frac{dW}{dy} = WF(y), \tag{12}$$

where $F(y) = \frac{9\sqrt{3}}{8\pi} y \int_y^{\infty} K_{\frac{5}{3}}(x)dx$, $K_p(x)$ is the Macdonald function, $y = \frac{\omega}{\omega_c}$, and $\omega_c = \omega_0 \gamma^3$.

The spectral power density of the radiation in the low frequency region ($y \ll 1$) is given by

$$\frac{dW}{d\omega} \simeq \frac{e^2 \omega_c}{c\varepsilon_0 \gamma^2} \left(\frac{\omega}{\omega_c}\right)^{\frac{1}{3}}.$$

In the high-frequency region ($y \gg 1$) the spectral power density has the form:

$$\frac{dW}{d\omega} \simeq \frac{e^2 \omega_c}{c\varepsilon_0 \gamma^2} \left(\frac{\omega}{\omega_c}\right)^{\frac{1}{2}} exp\left(-\frac{2}{3}\frac{\omega}{\omega_c}\right). \quad (13)$$

In Fig. 5 the spectral distribution of the radiation is presented. The spectral distribution has a maximum near $\omega \approx \frac{\omega_c}{3}$, where $\omega_c = (\frac{3}{2})\omega_0 \gamma^3$, and the main part of the radiation is concentrated in this frequency range. These radiation frequencies are very large compared to the distance between two adjacent ones. The radiation spectrum consists of a very large number of closely spaced lines, i.e. it has a quasi-continuous character.

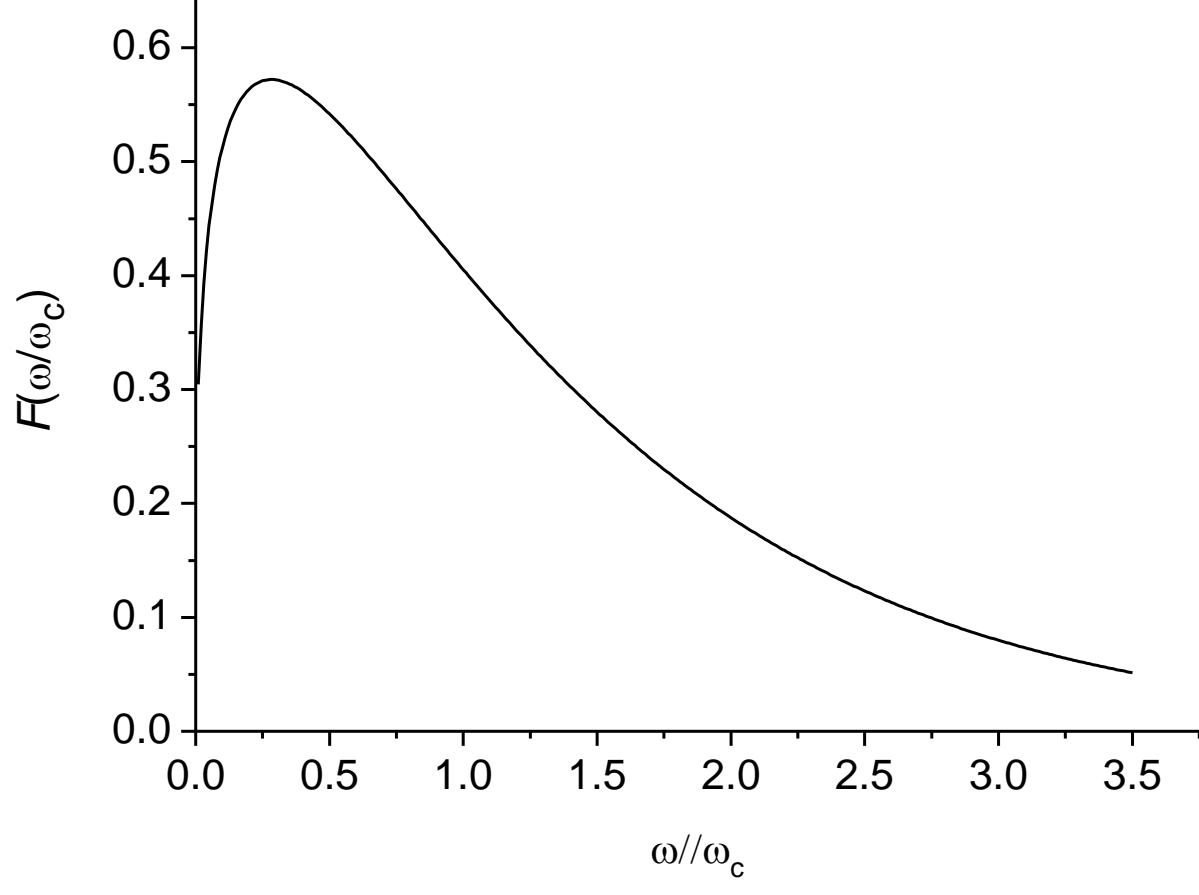

**Fig. 5.** Spectral distribution of the radiation.

The maximum in the spectral power distribution is achieved at the frequency $\omega_{max} = \frac{1}{2}\gamma^3 \omega_0$ , where $\omega_0 = \frac{v}{R}$ [37]. The radiation is concentrated mainly in a narrow cone with an axis along the direction of the velocity. The angular width $\Delta\theta$ in which the main part of the radiation is enclosed is inverse proportional to the relativistic Lorentz factor: $\Delta\theta \sim \frac{1}{\gamma}$. This indicates that the ultra-relativistic velocities of the polarization bunch lead to a very narrow spatial distribution of X-rays and gamma rays. A strong linear polarization of radiation occurs in the orbital plane. Moreover, the degree of polarization is equal to 3/4 [37]. Total powers of $\sigma$- and π- polarization components $W_\sigma$ and $W_\pi$, are equal to $W_\sigma = \frac{7}{8}$ and $W_\pi = \frac{1}{8}$ [36]. In the plane of the orbit of revolution, the radiation is completely linearly polarized, since $W_\pi = 0$. Note that polarization can serve as an accurate criterion for testing hypotheses about the nature of radiation. Based on the results of measurements of the polarization of electromagnetic radiation of the Crab Nebula, the synchrotron nature of the radiation was established [38].

The spectral power density (12) can be transformed into the energy distribution of the photon flux, which is the number of photons with energy $E_{ph}$ emitted per second ($\dot{N} = dN/dt$), into the energy band $dE_{ph}$:

$$dW = E_{ph} d\dot{N} = E_{ph} \frac{d\dot{N}}{dE_{ph}} dE_{ph},$$

$$\frac{dW}{d\omega} = \hbar E_{ph} \frac{d\dot{N}}{dE_{ph}}, \quad \frac{d\dot{N}}{dE_{ph}} = \frac{1}{\hbar E_{ph}} \frac{dW}{d\omega} = \frac{W}{E_{ph} E_c} F\left(\frac{\omega}{\omega_c}\right), \; E_c = \hbar\omega_c.$$

In Fig. 6 the calculated and measured photon counts $d\dot{N}$ per energy band $dE_{ph}$ are presented.

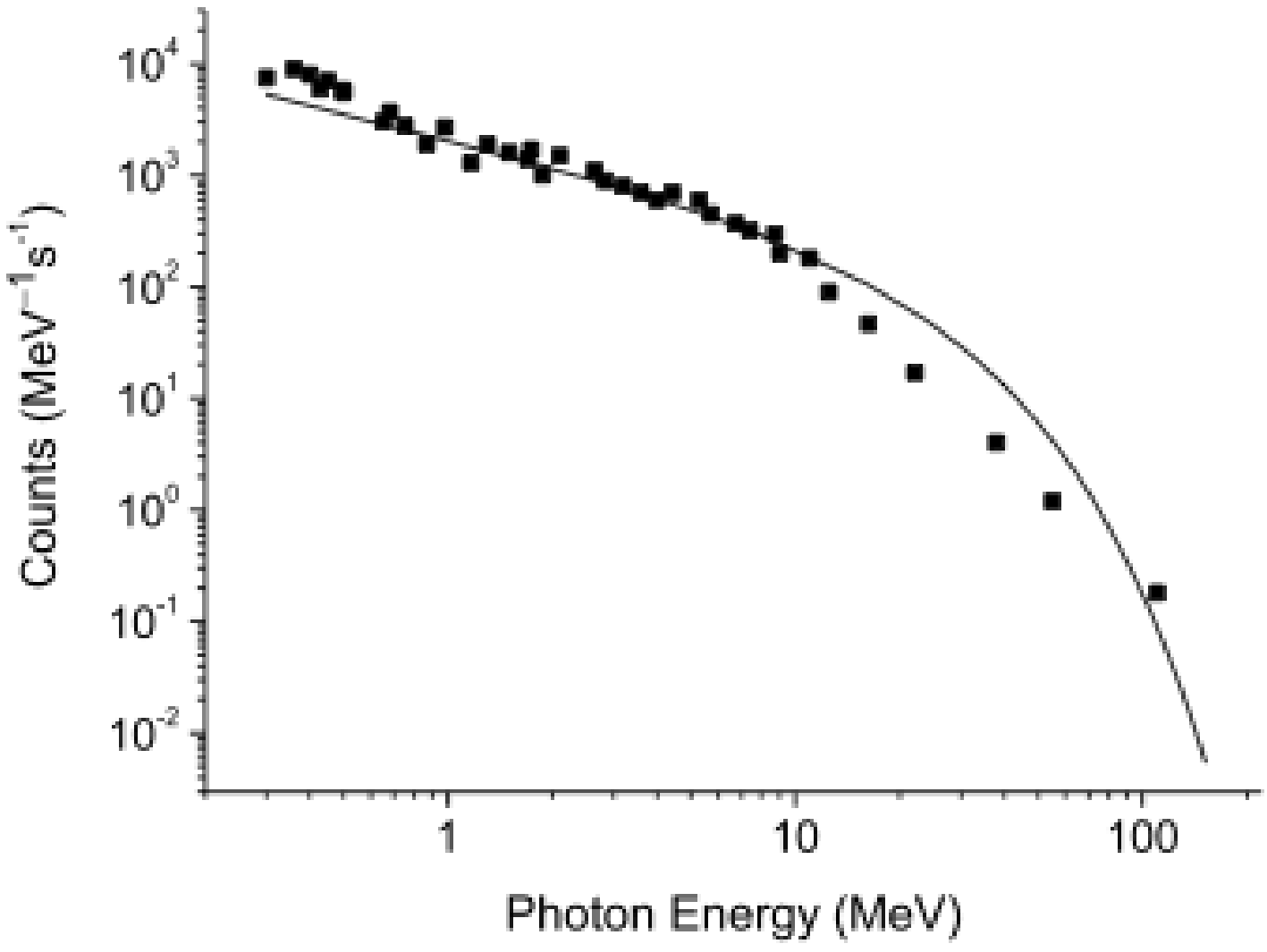

**Fig. 6.** The normalized photon spectrum (solid curve) and measured spectrum from AGILE [20]. $E_c$ = 16 MeV

It is seen that there is not exponential cutoff near 10 MeV in contrast to the RREA models.

Our model is also consistent with the measured spectrum in a laboratory spark discharge [15]. In Fig. 7 the calculated and measured photon counts are presented. A power-law spectrum is seen for energies less than 1 MeV, while the exponential decrease is observed at high energies.

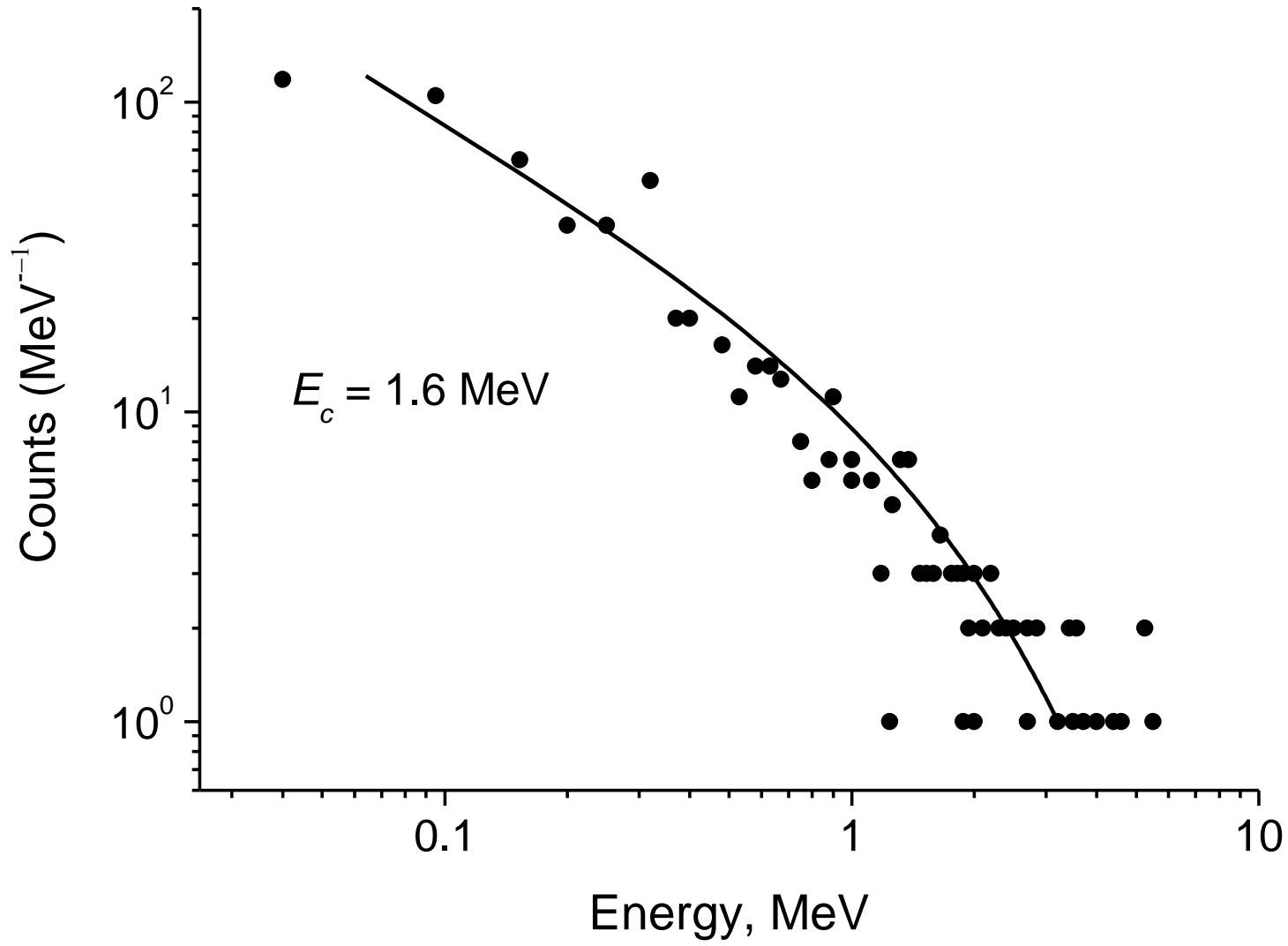

**Fig. 7.** The normalized photon spectrum (solid line) and measured spectrum from laboratory spark discharge [15]. $E_c = 1.6$ MeV.

The number of photons with frequencies $\omega_{max} \approx \omega_0 \gamma^3$ radiated for the time of the electron's revolution $\delta t \approx \frac{1}{\omega_0}$ around the circle is given by

$$N_{ph} \approx \frac{W}{\omega_0\, \hbar\omega_{max}} \approx \frac{e^2}{\hbar c \varepsilon_0} \gamma \approx 10^{-1}\gamma$$

For single charges, the radiation is weak. The situation changes when we consider bunches of particles. Indeed, the polarization of the plasma occurs in a certain extended region, so as a result, the amplitude of the polarization radiation will be determined by the total dipole moment of the bunch. In this case, a bunch of polarized plasma will radiate like point particles with a charge and multipole moments corresponding to the entire bunch.

The radiation in the high-frequency range is incoherent and proportional to the number of electrons [37]: $W_N = N_e \cdot W$.

### 3.2. Photon frequency

In Fig. 8 the photon frequency as function of the conductivity is presented for different frequencies of the surface wave.

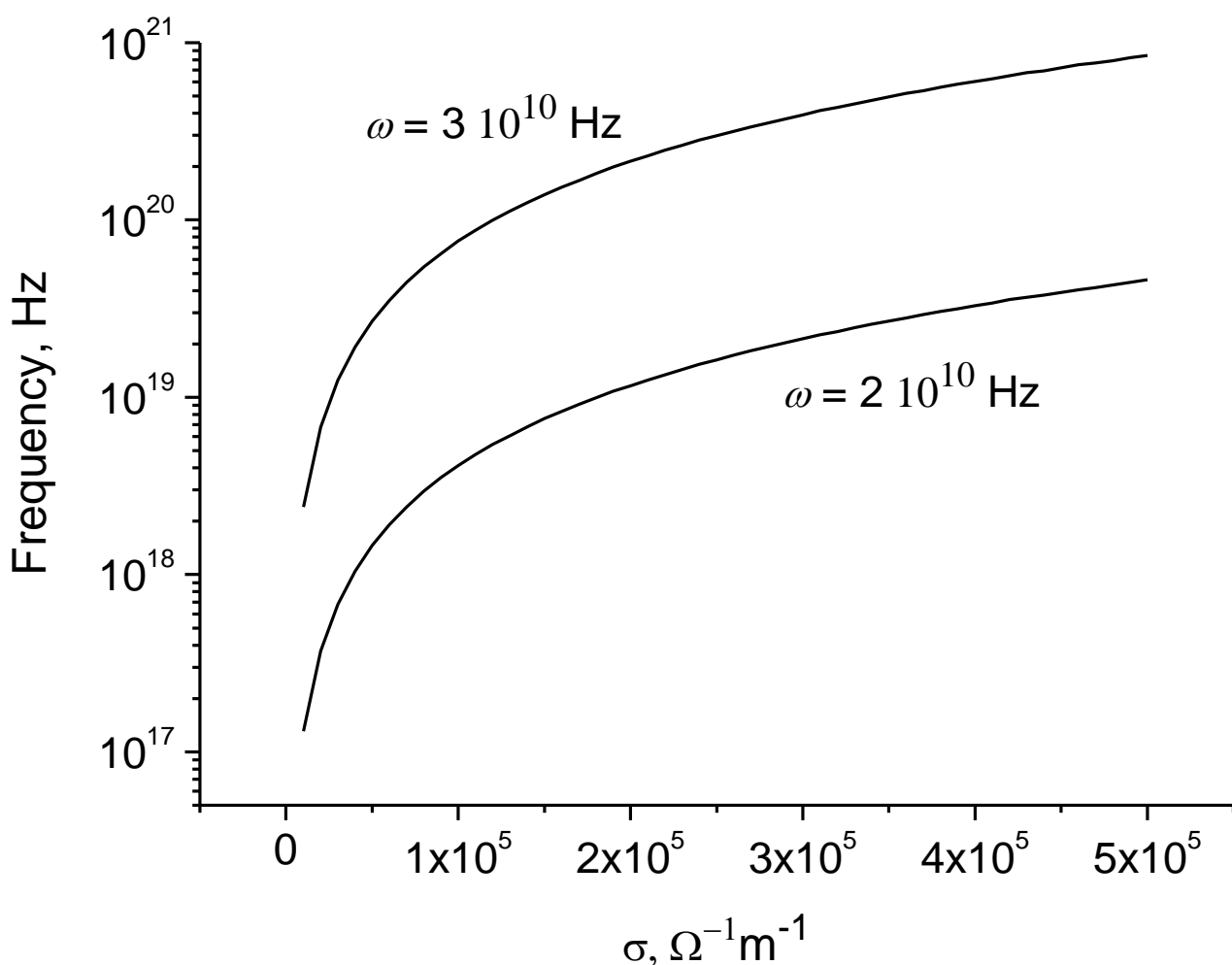

**Fig. 8.** Photon frequency as function of conductivity. $r_0$ = 0.01 m.

In Fig. 9a the values $\gamma = \left(1 - \frac{v^2}{c^2}\right)^{-\frac{1}{2}}$ as a function of the conductivity are presented. In Fig 9b the photon frequency as a function of the conductivity is shown.

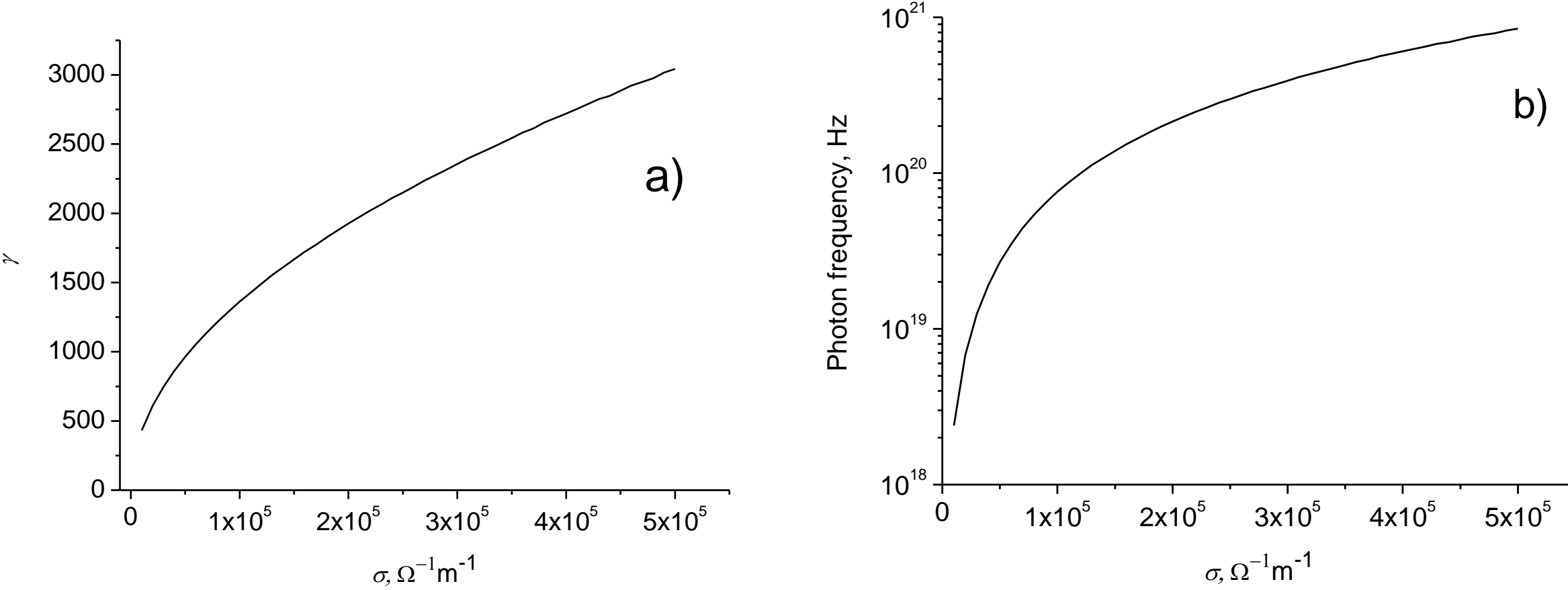

**Fig. 9.** The Lorentz factor (a) and photon frequency (b) as function of conductivity. $\omega = 2\pi f = 3 \cdot 10^{10}$ Hz, $r_0$ = 0.01 m.

Thus, the spectrum of synchrotron radiation of lightning covers almost the entire scale of electromagnetic waves - from the radio frequency to the X-rays and gamma rays.

In Table 1 the attenuation lengths, Lorentz factors, photon frequencies and energies are presented for different values of the channel conductivity and surface wave frequency.

Table 1. Relationships between lightning and X-ray and gamma-ray radiation parameters.

$r_0 = 0.01$ m, $f = \frac{\omega}{2\pi}$.

| $\sigma, \Omega^{-1}m^{-1}$ | $f$, Hz | $z_0$, m | $\gamma$ | $\omega_{ph}$, Hz | $E_{ph}$, eV |
|---|---|---|---|---|---|
| $10^4$ | $10^8$ | 175 | 17 | $1.48 \cdot 10^{14}$ | 0.1 |
| | $10^9$ | 27.2 | 27.7 | $0.64 \cdot 10^{15}$ | 0.42 |
| | $5 \cdot 10^9$ | 4.2 | 441 | $2.57 \cdot 10^{18}$ | 1690 |
| $10^5$ | $10^8$ | 556 | 31.7 | $0.96 \cdot 10^{15}$ | 0.63 |
| | $10^9$ | 85.5 | 52.8 | $4.4 \cdot 10^{15}$ | 2.9 |
| | $5 \cdot 10^9$ | 13.3 | 1393 | $8.1 \cdot 10^{19}$ | 53350 |
| $10^6$ | $10^8$ | 1667 | 59 | $6.16 \cdot 10^{15}$ | 4.05 |
| | $10^9$ | 267 | 100 | $3 \cdot 10^{16}$ | 19.7 |
| | $5 \cdot 10^9$ | 42 | 4389 | $2.54 \cdot 10^{21}$ | 1670000 |

### 3.3. Photon number

The number of photons with frequencies $\omega_{max} \approx \omega_0 \gamma^3$ radiated for the time $\delta t \approx \frac{1}{\omega_0}$ of the electron's revolution around the circle is given by

$$N_{ph} \approx \frac{W}{\omega_0\, \hbar\omega_{max}} \approx \frac{e^2}{\hbar c \varepsilon_0}\gamma \approx \frac{2.56 \cdot 10^{-38}\gamma}{1.05 \cdot 10^{-34} \cdot 3 \cdot 10^8 \cdot 8.85 \cdot 10^{-12}} \approx 10^{-1}\gamma. \tag{14}$$

This number of photons corresponds to the radiation over time $\delta t \sim 10^{-10}$ s. For a time of 1 μs, the number of photons is equal to $N_{ph} \sim 10^3\gamma$.

The power of synchrotron radiation consists of coherent and incoherent parts [37]:

$$W(m) = W^{incoh}(m) + W^{coh}(m).$$

When radiation is incoherent, the total power is given by [37]:

$$W_N(m) = N_e \cdot W(m),$$

where $N_e$ is the number of electrons in a polarized plasma bunch.

Number of electrons in a polarized plasma bunch is determined by the polarization current $i_{pol}$:

$$N_e \simeq \frac{Q}{e} \simeq \frac{i_{pol} \cdot \Delta t}{e},$$

where $\Delta t \sim \tau_f \sim \frac{1}{\omega} \sim 10^{-10} s$.

For the electrons number we obtain

$$N_e \simeq \frac{Q}{e} \simeq \frac{i_{pol} \cdot \Delta t}{e} \sim \frac{10^3 \cdot 10^{-9}}{1.6 \cdot 10^{-19}} \sim 10^{12 \div 13}.$$

Thus, for the number of photons radiated for a time of 1 μs we have:

$$N_{ph} \approx 10^3 \gamma \cdot N_e \approx 10^3 \cdot 10^3 \cdot 10^{12\div13} \sim 10^{18\div19} \text{ photons.}$$

This estimated photon number is in the range of observation data. Note that the typical brightness of a TGF observed from space is within an order of magnitude of $10^{17}$ - $10^{19}$ gamma rays [17, 39].

The radiation power increases dramatically in the case of coherent bunch of electrons grouped at distances less than the wavelength of the emitted wave. The coherent radiation can be observed in low-frequency range, i.e., in the radiofrequency one. It is known that the radio-frequency pulses of pulsars are emitted by a coherent synchrotron mechanism [40]. The radiation power increases dramatically in the case of coherent bunch of electrons grouped at the distances less than the wavelength of the emitted wave. In this case the radiation power is given by $W^{tot} = N_e^2 \cdot W$.

High-intensity microwave radiation of a lightning discharge was observed in [41-43]. The extreme brightness of the radiation observed from fast radio bursts indicates that it is generated in the process of coherent radiation. It was shown in [42] that the microwave radiation from lightning is a sequence of individual pulses, and the radiation spectrum differs from the spectrum in the long-wave range. This indicates that in the decimeter range, the mechanism of generating electromagnetic waves differs from the usual dipole radiation of the lightning current. Recently it was shown that each leader step emits a burst of multiple discrete VHF pulses [22].

## 4. Fine structure of lightning and spark discharge channel

The trajectory of a lightning leader has a complex 3D structure in which the tortuous main channel and branches deviate from the electrical force lines. Tortuosity and branching is one of the main peculiarities of the lightning discharge. In a real lightning environment, the tortuosity and branching of the lightning channel make a significant contribution to the probability of lightning strikes to complex objects [44-48]. A powerful fractal approach can be used to model the orientation of lightning to structures considering the random behaviour of lightning trajectory [44-47]. The large-scale tortuosity in the lightning channel is usually of the order of the streamer zone length of leaders, which is around 10-100 m. If the tortuosity and branching corresponding to the characteristic length of the order of 10-100 m can be clearly seen using conventional devices such as cameras, then the fine structure of the order of the transverse size of the lightning channel was very rarely observed. However, the fine structures of lightning channel can be also detected at close distances and sufficient camera resolution. In [49] the diameter of the lightning stroke was measured by allowing lightning discharges to pass through fiberglass screen. The diameters of the holes produced in fiberglass screen were varied from 2 mm to 3.5 cm. It was shown in [50] that the channel diameter determined from the 224 images obtained with a high-speed framing camera has a mean value of 6.5 cm. In a laboratory spark the diameter of channel lies in the range of 2-4 mm, although the photos show the diameters of the order of 3-5 cm. Diameters of lightning channels from photos lies in the range 10-45 cm. This indicates that the current cord diameter is an order of magnitude less than the diameters determined from conventional images. Therefore, high resolution images taken at different stages of the return stroke are important in order to resolve the fine structure of lightning channel.

Usually, photos of lightning flashes are taken with excessive lighting and insufficient resolution, which is why small details of the breakdown channel are not visible. However, from high-speed photographs, it is possible to select temporary sequences of lightning images on which the tortuosity of the breakdown channel can be clearly visualized on a centimeter and decimeter scale. Below some examples of irregular and regular spiral forms of spark and lightning channels are presented.

In Fig. 10 an irregular spiral structure of a positive polarity spark channel at the return stroke phase in a long air gap of 6 m is shown [51]. Surprisingly, the spark channel does not pass through the metal sphere because of the repulsion effect of positive leader from the metal sphere [52].

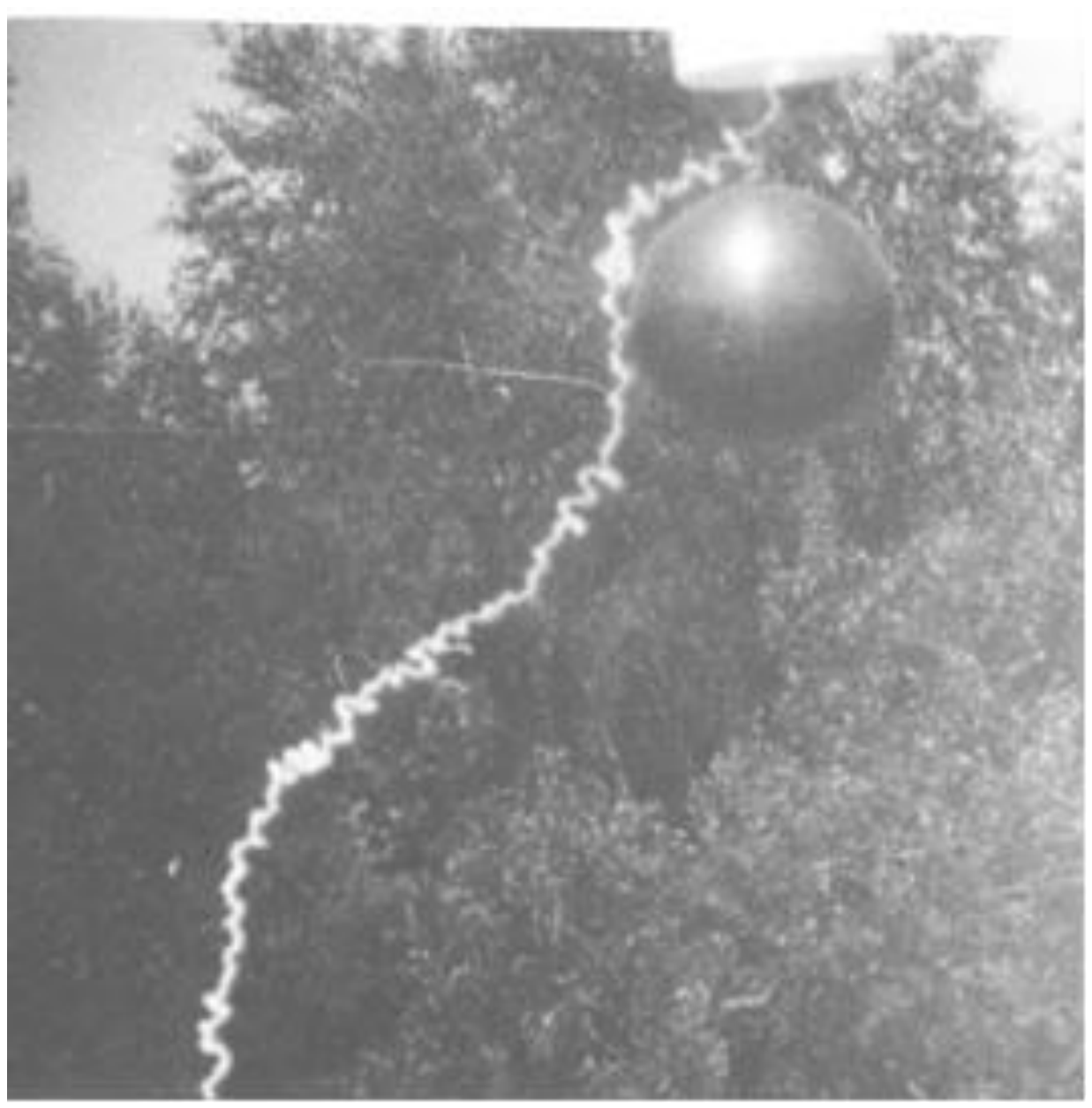

**Fig. 10.** Vortex current trajectory in a laboratory positive spark discharge in a 6 m long air gap (adapted from [51]). The step of spiral is equal to 1.5÷2.5 cm, diameter of spiral $\phi \approx 3$ cm, diameter of metal sphere $D_{sp} = 30$ cm.

In this case, the centimeter-scale tortuosity of the spark channel is formed not at the stage of leader propagation, but at the return stroke phase. This indicates that a return stroke current path with a small-scale tortuosity is created around the original leader channel, which does not have a small-scale tortuosity.

In Fig. 11 one of the recordings of the decimeter-scale tortuosity of the lightning channel is shown [53]. It is most likely that the photography illustrates the portion of the lightning channel corresponding to the upward positive connecting leader.

It is seen that unlike the laboratory spark discharge channel, the regular spiral parts are seen in the photo (Fig. 11a). If the length of the channel is about 10 m, then the spiral pitch will be of the order of $h$ = 24 cm and the spiral radius of the order of 12 cm.

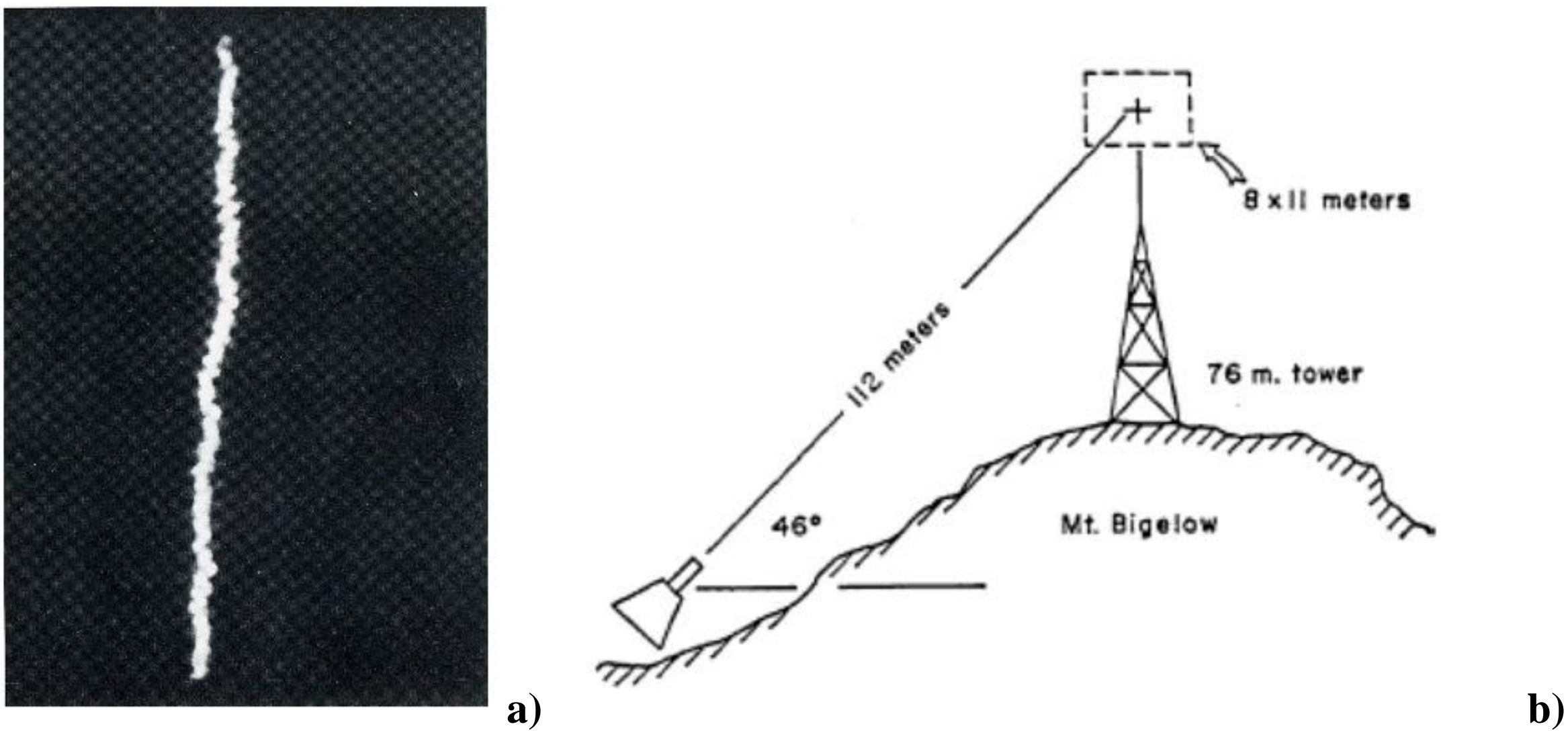

**Fig. 11.** (a) Fine scale tortuosity observed in the bottom 10-15 m of the lightning flash to a tower of height 74 m located on a mountain. (b) Geometry of a tower location (adapted from [53]).

In Fig. 12 the photo of a small section of the lightning channel, which was taken with an ordinary camera, is shown [54]. The negative image shows three moments of lightning's life. In the center of the image, the initial stage of the lightning's life is fixed. The last stage of lightning's life is shown below. The upper part of the photo shows the intermediate stage of the development of lightning, in which the spiral structure is clearly visible. Taking the diameter of the lightning channel ~ 0.2 m, it is possible to determine the spiral pitch $h$ ~ 0.1 m from the photo. Here, the counterclockwise helicity is clearly seen.

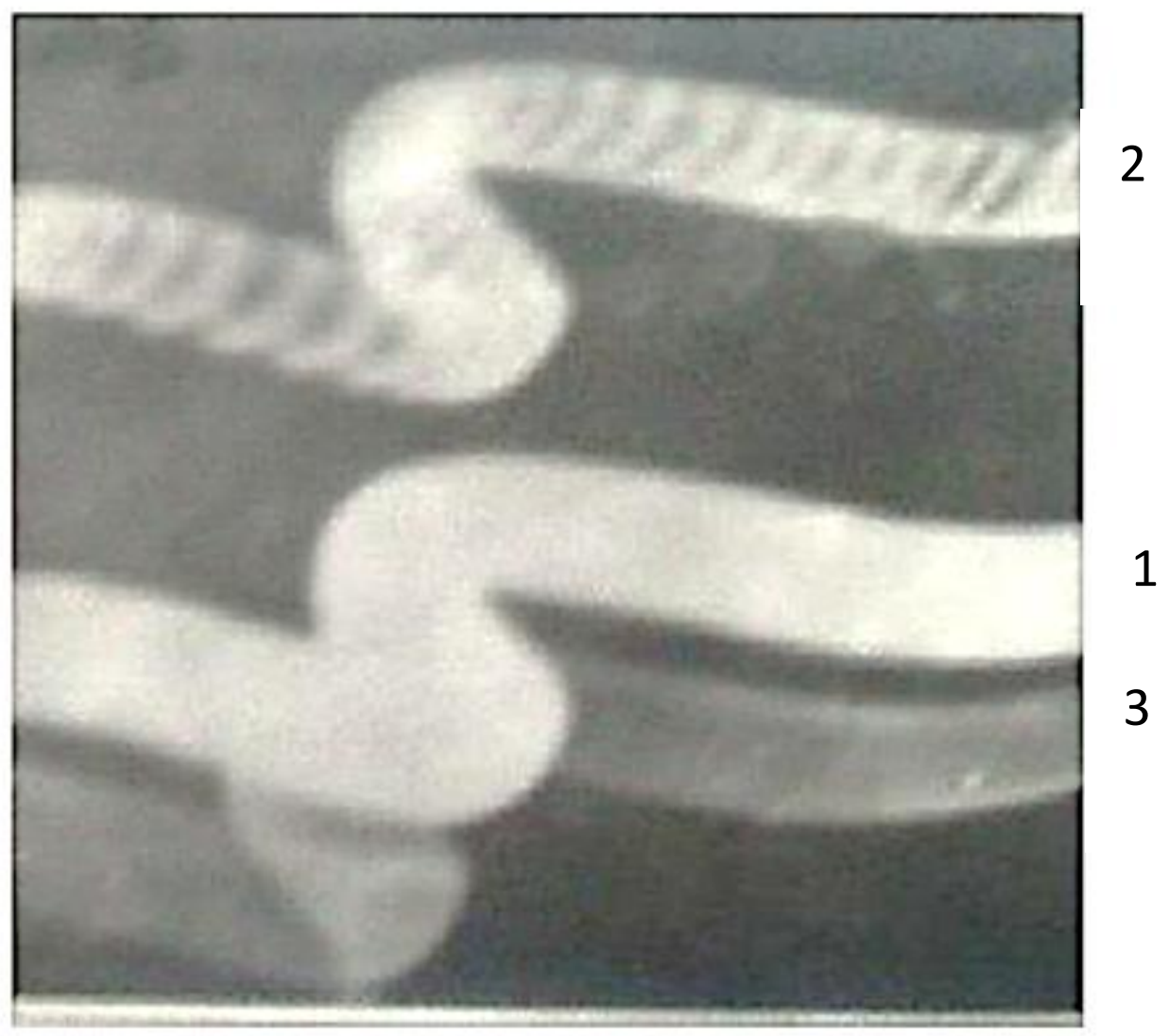

**Fig. 12.** Photography by enlarging a portion of the image of the lightning channel: $\emptyset = 20\ cm$; $h = 10\ cm$. (adapted from [54]). 1 – initial stage, 2 – intermediate stage, 3 – last stage.

It is seen that the radius of the cyclotron motion is practically constant. This indicates that the centripetal force $F = ma_r = mv^2/R$ is compensated by the Lorentz force $F_L = evB_{eff}$. Therefore, it is possible to estimate the magnitude of the magnetic field created by a vortex current.

$$B = \frac{mv}{eR} \approx \frac{10^{-30} \cdot 2 \cdot 10^{8}}{1.6 \cdot 10^{-19} \cdot 10^{-1}} \sim\ 10^{-2}\ T\ . \qquad (15)$$

On the other hand, the axial magnetic field created by a vortex current $i_v$ is expressed by

$$B = \mu_0 \frac{i_v}{2R}. \qquad (16)$$

Using (15) and (16) we obtain for the magnitude of the vortex current

$$i_v = \frac{2BR}{\mu_0} \approx \frac{2 \cdot 10^{-2} \cdot 10^{-1}}{1.3 \cdot 10^{-6}} \sim 1.5 \cdot 10^{3}\ \text{A}.$$

Since the radius of the spiral in Fig. 12 is about 0.1 m, it follows that the observed spiral trajectory corresponds to a current value of the order of several kA. Note that the amplitude of the return stroke current can reach values of the order of 100 kA [4, 30, 31], i.e. the vortex current is significantly less than the peak values of the return stroke current.

Figure 13 shows an enlarged portion of the lightning flash channel, recorded by Charles Moussette in 1886 [55]. This photo clearly demonstrates the irregular spiral shape of the lightning channel.

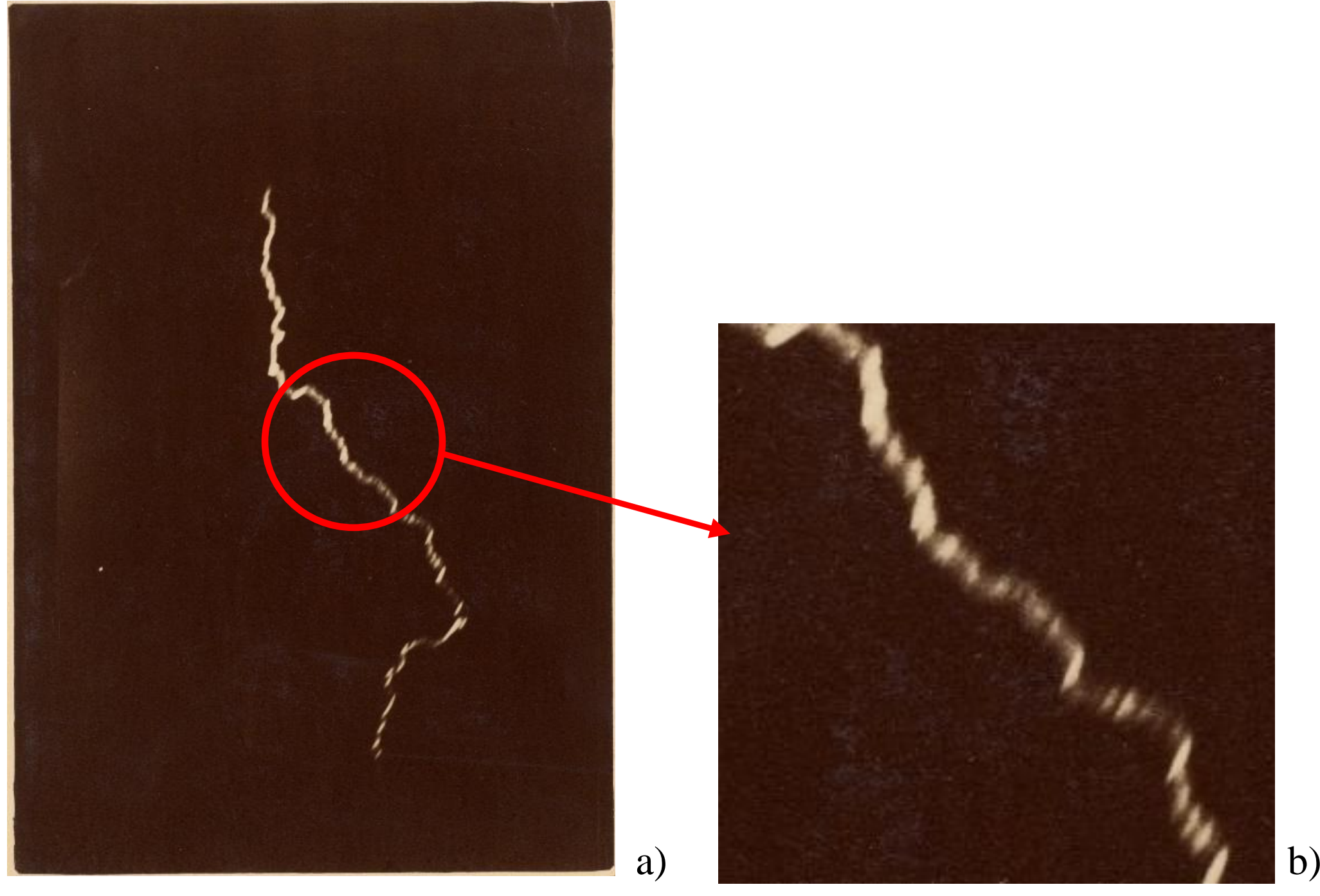

**Fig. 13.** (a) Irregular spiral form enlarged image of lightning in a thunderstorm (adapted from [55]), (b) Enlarged image of the central part of (a).

The photos presented clearly demonstrate centimeter- and decimeter-scale tortuosity of spark and lightning channels. Note that this small-scale channel geometry is usually missing in the leader propagation stage. This means that a small-scale spiral channel is not created by the leader, but during the neutralization of charges introduced by the leader-streamer system, i.e. when a high-intensity ionization wave propagates through previously created leader channels. This may occur during the final jump phase – the very beginning of the first and subsequent return stroke stages.

The reason of the formation of the charge neutralization vortex current outside the preliminary established leader channel is the skin effect, which causes strong dissipation. The waves cannot penetrate the internal part of the channel. Besides, the resistance of the channel increases with frequency. The skin layer thickness obtained from the consideration of the current density is given by

$\delta = \frac{1}{\sqrt{2\pi}}\sqrt{\frac{\varepsilon_0}{\sigma\omega}}$ , where $\sigma$ is the electric conductivity, $\omega$ is the frequency. Because of the current density changes with the radius proportionally to $e^{-(r_0-r)/\delta}$, the interior of the channel is almost free of currents at high frequencies.

## 5. Discussion

Thus, we have shown the existence of fast electromagnetic surface plasmon waves (SPW) propagating along the lightning discharge channel at a relativistic speed close to the speed of light in vacuum. It follows from our simulations, that the higher the electrical conductivity of the channel and the frequency of SPW, the higher the speed of the polarization current pulse. The propagation distance of the surface wave decreases with increasing frequency.

At relativistic speeds the spin properties of electrons along with a charge become significant. The movement of electrons considering their spin in an electric field created by a plasma lightning channel, leads to the appearance of an effective magnetic field due to the quantum spin Hall effect caused by the relativistic spin-orbit interaction phenomenon [56-59]. As a result, a vortex current flow is formed along the lightning channel in the absence of an external magnetic field. The proposed theoretical model of the formation of a spin-induced vortex current in a plasma channel is consistent with observations of lightning and spark discharges. It is shown that the current pulse moving along the spiral trajectory with relativistic speed can produce microwave, THz, X-ray and gamma-ray emissions via cyclotron and synchrotron radiation mechanisms. Conventionally, X-ray and γ-ray emissions, observed in correlation with negative leaders of lightning and long sparks of laboratory experiments, are associated with the bremsstrahlung of high-energy runaway electrons [60]. Here we have proposed a synchrotron mechanism of X-ray and gamma-ray emissions, when the ionization wave (SPW) of the return stroke propagates along a spiral trajectory formed around the ionized channel previously created by the leader. SPWs are the fast ionization waves that are generated during the lightning leader stepwise propagation, final jump phase – the very beginning of the first and subsequent return stroke stages. The fast high intensity ionization waves can be also generated during the neutralization of charges injected into the space by a dart leader and a recoil leader [61]. Under such conditions, we calculated the radiation frequencies as a function of the polarization current velocity (which depends on the conductivity and radius of the channel) and explained some basic properties, such as the radiation spectrum (in the region above 10 MeV), the cut-off energy, as well as the polarization and beaming characteristics of the radiation. We found that it is possible to observe radiation with a photon spectrum of up to 5 MeV in spark discharges and up to 100 MeV in lightning. In [62], electromagnetic pulses of spark discharge radiation with a front duration of 50-100 ps and a pulse duration of 1.5 ns in the frequency band up to 10 GHz were recorded. Linear polarization of the radiation field was observed, which is in good agreement with the cyclotron mechanism.

One of the conditions for confirming the synchrotron radiation mechanism is the movement of charges along a spiral trajectory. Such a movement of charges is possible in the presence of a magnetic field directed along the axis of the breakdown channel. We have shown that an effective magnetic field arises due to the spin properties of electrons. It should be noted that consideration the spin properties of electrons becomes important when relativistic effects become significant, i.e. when the propagation velocities of a current pulse practically reach the speed of light. For the appearance of a regular periodic spiral structure of the breakdown channel, it is necessary that the Lorentz force compensates for Coulomb forces and diffusive pressure forces, which is fulfilled with high channel conductivity. These conditions are fulfilled in the lightning channel. In a spark discharge, the conductivity and concentration of electrons are less than in lightning, and the propagation velocity of the surface plasmon wave is noticeably less than the speed of light, which is manifested in irregular small-scale channel geometry.

The formation of small-scale tortuosity in the laboratory discharge may also occur due to instability in the leader channel. Indeed, the laboratory gas discharges also contain an eddy electric current, which is associated with the noncollinearity of the gradients of plasma density and electrons temperature [63]. The density of the eddy current is an order of magnitude less than the discharge current, so when the main current is added, it modulates its radial profile slightly. In a stratified discharge in a uniform magnetic field, an eddy electric current due to the noncollinearity of the plasma-density and temperature gradients brings the gas to rotation [64].

Note that the tortuosity of a channel in centimeter and decimeter scales can be observed in laboratory sparks and natural lightning. However, the difficulties of registration due to excessive illumination of the image and the need to increase the resolution do not yet allow to study in detail the features of channels of small scale. Indeed, it was shown in [65] that the vhf-emitting width of the recoil leaders is thinner than the LOFAR meter-scale resolution.

We believe that the tortuosity of the channels on the centimeter and decimeter scales is a common feature of lightning and spark discharges. This is evidenced by high-resolution photographic data taken at close distances. Studies of the fine structure of the channel geometry will shed light on understanding the processes of neutralization of charges injected by streamer-leader system, including dart-leader and recoil-leader, as well as the mechanism of formation of subsequent strokes.

## 6. Conclusion

In summary, the propagation of SPP waves with ultra-relativistic speed along a lightning channel is demonstrated. It is emphasized that the spin properties of electron density current become important at relativistic velocities. A physical model of vortex current formation in a plasma channel of lightning discharges is proposed. It is shown that the spin-orbit coupling occurs as an effective magnetic force acting on the electron spin, leading to the formation of a spin-induced vortex current flow in the absence of an external magnetic field. The existence of long tails in the power spectrum of radiation is shown, which explains observations of photon energies in the range of 10-100 MeV in the TGF. The source of synchrotron radiation of lightning is the surface electromagnetic wave (surface plasmon-polariton), moving along a regular and irregular spiral ionized channel with a relativistic velocity. The unique properties of synchrotron radiation are a sharp angular directivity, strong linear polarization, and a wide spectrum with a maximum in the high frequency region.

Our simulations show that the theoretical model of X-ray and gamma-ray emissions, as well as radio-frequency radiation produced by Lightning discharge by a current pulse moving along a helical trajectory is consistent with the observational data. This indicates that the lightning discharge channel can be considered as an accelerator of charged particles producing microwave, X-ray and gamma-ray beams via cyclotron and synchrotron radiation mechanisms.

**Funding:** Ministry of Science and Higher Education of the Russian Federation (FFNS-2022-0009).